\documentclass[aps,superscriptaddress,twocolumn,a4paper,prl,showpacs]{revtex4-1}
\usepackage{latexsym,graphicx,amsmath,amsfonts}
\def\ket#1{| #1 \rangle}
\def\bra#1{\langle #1 |}
\def\proj#1{\ket{#1}\bra{#1}}
\def\tr{\text{Tr}}

\newcommand{\one}{\mathbb{I}}

\begin{document}
\title{Unitary Holonomies by Direct Degenerate Projections}

\author{Daniel~K.~L.~\surname{Oi}} \email{daniel.oi@strath.ac.uk}
\affiliation{SUPA Department of Physics, University of Strathclyde,
  Glasgow G4 0NG, United Kingdom}

\begin{abstract}
  An incomplete quantum measurement can induce non-trivial dynamics
  between degenerate subspaces, a closed sequence of such projections
  produces a non-abelian holonomy. We show how to induce unitary
  evolution on an initial subspace from such finite discrete sequences
  and also construct a near deterministic repeat-until-success
  protocol. We also prove necessary and sufficient criteria on the
  auxiliary Hilbert space dimension required for inducing isometries
  between between subspaces.
\end{abstract}

\date{\today}


\maketitle


The geometric structure of quantum theory is highlighted by the
phenomenon of the Pancharatnam-Berry Phase whereby the cyclic
evolution of a pure quantum state induces a geometric phase ($U(1)$
abelian holonomy) in addition to the standard dynamic
phase~\cite{Pancharatnam,Berry}. Non-abelian holonomies can be induced
by the cyclic adiabatic modulation of a Hamiltonian with a degenerate
subspace~\cite{WilczekZee1984} or by non-adiabatic
means~\cite{Anandan1988}. Alternatively, the evolution of a subspace
can be driven deterministically by a dense sequence of incomplete
(degenerate) projections again leading to a
holonomy~\cite{AP1989}. Such Zeno effects have been proposed for
quantum control and computation~\cite{Burgarth2013,Lidar2013}, and for
engineered quantum systems~\cite{Raimond2013}.

The more practical case of finite projective sequences was addressed
by Anandan and Pines~\cite{AP1989} and later by {\AA}berg, Kult, and
Sj{\"o}qvist~\cite{KAS2006,SKA2006} where they analysed the geometric
structure of sequences of points in the Stiefel manifold of projective
subspaces, and found the associated holonomies. Here, we extend this
by explicitly constructing finite discrete sequences of degenerate
projections that induce isometries between subspaces and demonstrate
two methods of achieving unitary holonomic evolution. The first is
minimal in that only one auxiliary Hilbert space dimension is
required, though at the expense of the success probability. The second
provides a near deterministic protocol, but requires a doubling of the
Hilbert space dimension. We also prove that this doubling is a
necessary condition for stepwise unitary (isometric) subspace dynamics.


In an $N$-dimensional Hilbert space we can perform an incomplete
measurement where one outcome is a degenerate projection onto a
$k$-dimensional subspace and the complementary result can be taken as
a projection onto a $N-k$ dimensional subspace. Without loss of
generality we will identify a projection operator with its $+1$
eigenspace or a set of basis vectors. Consider an initial state
$\ket{\psi_0}$ lying in a $k$-dimensional subspace associated with a
projector $\Pi_0$. Applying a second $k$-dimensional
projector $\Pi_1$ (assumed to be non-orthogonal to
$\Pi_0$), the system survives with probability $p_1=\bra{\psi_0}\Pi_1
\ket{\psi_0}$ and now lies within the subspace of $\Pi_1$. The
normalized conditional state is given by $\Pi_1
\ket{\psi_0}/\sqrt{p_1}$.

Extending this to a sequence of projections
$\{\Pi_j\}_{j=0}^{n}$ where the final projection $\Pi_n$ coincides
with $\Pi_0$, the system may undergo a cyclic evolution and return to
its original subspace. The final conditional state is related to the
initial state by
\begin{equation}
\ket{\psi_f}=\Gamma \ket{\psi_0}/\sqrt{p_{f}},
\end{equation}
where the cumulative operation is given by
\begin{equation}
\Gamma=\prod_j \Pi_j
\end{equation}
and the survival probability is $p_{f}=\bra{\psi_0}\Gamma^\dagger
\Gamma\ket{\psi_0}$.  In general, $\Gamma$ is not proportional to a
unitary operation on the initial subspace. In the limit of a dense
sequence of projections approaching a continuous path in the
associated Grassmann manifold, then $\Gamma$ becomes
unitary~\cite{KAS2006}.



\begin{figure}
\includegraphics[width=\columnwidth]{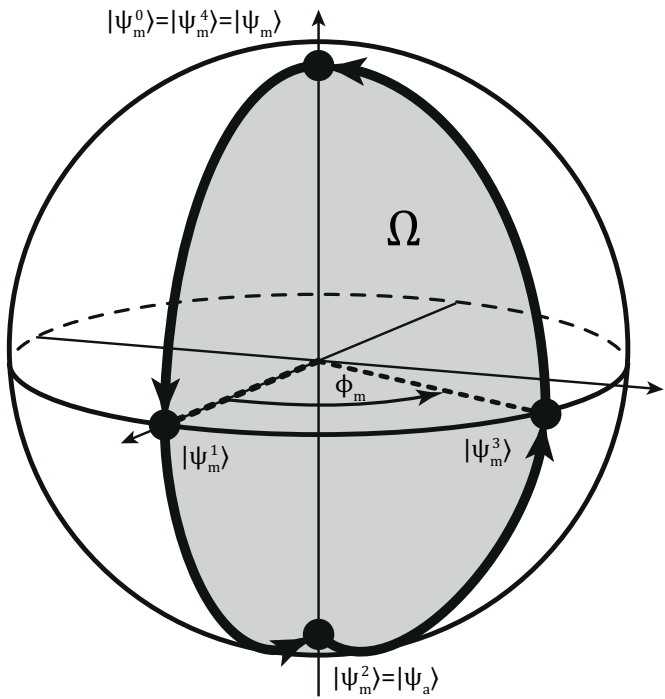}
\caption{Geometric phase of the $\ket{\psi_m}$ component. The
  $\ket{\psi_m}$ component of the $\Pi_0$ subspace is driven by a
  projective subsequence along the closed path indicated by the thick
  arrows within a two dimensional subspace represented by a Bloch
  sphere. The transition amplitudes are equal in magnitude between the
  four points (solid circles). The Pancharatnam or geometric phase
  associated with the cyclic evolution is half of the solid angle
  $\Omega$ enclosed (shaded) and is equal to $\phi_m$.}
\label{fig:bloch}
\end{figure}

We require only one additional dimension ($N=k+1$) in order to
generate a unitarily proportional $\Gamma$ using a finite sequence.
To illustrate, we construct a unitary operation $U=\sum
e^{i\phi_m}\proj{\psi_m}$ chosen to be diagonal in some orthonormal
basis $\{\ket{\psi_m}\}_{m=1}^k$ for the initial subspace. The process
proceeds stepwise by generating each phase factor in turn by a
sub-sequence of projections driving the $m^{\text{th}}$ component of
the superposition around a loop within a two-dimensional subspace
spanned by $\ket{\psi_m}$ and a single auxiliary level $\ket{\psi_a}$.
Each loop generates a geometric phase equal to half of the solid angle
enclosed on the Bloch sphere.

Specifically, to generate $\phi_m$ we use a sub-sequence
$\{\Pi_m^l\}_{l=0}^{N_m}$ where the starting and ending subspaces
coincides with $\Pi_0$, i.e. $\Pi_m^0=\Pi_m^{N_m}=\Pi_0$,
\begin{equation}
\label{eq:singlephase}
\Pi_m^l=\left(\sum_{j\ne m}
  \proj{\psi_j}\right)+\proj{\psi_m^l},
\end{equation}
with $\ket{\psi_m^0}=\ket{\psi_m^{N_m}}=\ket{\psi_m}$, and
$\ket{\psi_m^l}$ traces a path in the the subspace of
$\{\ket{\psi_m},\ket{\psi_a}\}$. The first term on the right hand side
simply projects onto all but one of the spanning basis vectors on
$\Pi_0$. The cyclic evolution of the remaining term is responsible for
generating the $\phi_m$-phase (Fig.~\ref{fig:bloch}). An initial state
$\ket{\psi}=\sum \alpha_j \ket{\psi_j}$ in the image of $\Pi_0$ will
transform under the sequence $\Gamma_m=\prod_l \Pi_m^l$ into the
unnormalized state $\sum_{j\ne m} \alpha_j \ket{\psi_j} + t_m \alpha_m
\ket{\psi_m}$ where
$t_m=\tr\left[\prod_{l=0}^{N_m}\proj{\psi_m^l}\right]$ and $\arg
t_m=\phi_m$. We require that $|t_m|>0$ for a non-trivial success
probability. A possible subsequence $\Pi_m^l$ can be specified by
\begin{eqnarray}
  \ket{\psi_m^0}&=&\ket{\psi_m^4}=\ket{\psi_m},
  \quad\ket{\psi_m^2}=\ket{\psi_a},\nonumber\\
  \ket{\psi_m^1}&=&\frac{1}{\sqrt{2}}
  \left(\ket{\psi_m}+\ket{\psi_a}\right),\nonumber\\
  \ket{\psi_m^3}&=&\frac{1}{\sqrt{2}}
  \left(\ket{\psi_m}+e^{i\phi_m}\ket{\psi_a}\right),
\end{eqnarray}
with $|t_m|^2=\left(\frac{1}{2}\right)^4=\frac{1}{16}$. Increasing
$N_m$ would enable the transition probability to increase until in the
limit of $N_m\rightarrow\infty$ we induce the Zeno effect and
$t_m\rightarrow 1$.

Applying $\Gamma_m$ for each $m$ leads to the final unnormalized state
\begin{equation}
\ket{\psi'}=\sum_m t_m \alpha_m \ket{\psi_m}=\Gamma\ket{\psi},
\end{equation}
where $\Gamma=\prod_m \Gamma_m=\sum t_m\proj{\psi_m}$. For the
conditional operation to be unitary on the initial subspace, all the
amplitudes should be reduced by the same factor so that the survival
probability is independent of the initial state, i.e. $|t_m|=t$. This
can be ensured either through suitable choice of $\Gamma_m$, or by a
final filtering operation to equalize $t_m$ to their smallest
magnitude. The success probability of any initial state is then $t^2$
and the conditional evolution is unitary as required,
$\ket{\psi'}/\|\ket{\psi'}\|=(\Gamma/t)\ket{\psi}=U\ket{\psi}$.


The procedure above creates a final unitary operation from the
conditional success of several non-unitary steps. Any information gain
at each step is offset in subsequent projections so overall no
information is gained about $\ket{\psi}$ conditional on all steps
succeeding. An alternate procedure would ensure that every transition
is an isometry between the source and image subspaces
and this requirement leads to a restriction on the minimum dimension
$N$ of the embedding Hilbert space, as shown below.


Let $\Pi_0$ and $\Pi_1$ be nonorthogonal $k$-degenerate projectors
specified with orthonormal bases
$\mathcal{B}_0=\left\{\ket{\mu_j}\right\}_{j=1}^k$ and
$\mathcal{B}_1=\left\{\ket{\nu_j}\right\}_{j=1}^k$ respectively. Let
$\text{span}(\mathcal{B}_0\bigcup\mathcal{B}_1)$ be
$(k+k')$-dimensional, $1\le k' \le k$. We can augment $\mathcal{B}_0$
with $k'$ extra vectors $\{\ket{\mu_j}\}_{j=k+1}^{k+k'}$ to form a
basis $\mathcal{B}_0^{'}=\left\{\ket{\mu_j}\right\}_{j=1}^{k+k'}$ for
the combined subspace. We now use the augmented basis
$\mathcal{B}_0^{'}$ to express the vectors of $\mathcal{B}_1$ as
$\{\ket{\nu_{j'}}=\sum_{j=1}^{k+k'} C_{j'j}\ket{\mu_j}\}_{j'=1}^{k}$,
where $C$ is a $k\times(k+k')$ complex matrix with orthonormal
rows. Using Gaussian elimination on $C$, we can find a matrix $D$
which defines a new orthonormal basis for $\Pi_1$,
$\mathcal{B}_1^{'}=\{\ket{\nu_{j'}^{'}}=\sum_{j=1}^{k+k'}
D_{j'j}\ket{\mu_j}\}_{j'=1}^k$, where
$\{\ket{\nu_{j'}^{'}}\}_{j'=1}^{k-k'}\subset\text{span}(\mathcal{B}_0)$,
and only $\left\{\ket{\nu_{j'}'}\right\}_{l'=k-k'+1}^k$ have support
outside of $\text{span}(\mathcal{B}_0)$. This implies that $\Pi_0$ and
$\Pi_1$ share a $(k-k')$ dimensional subspace. The $k'$ elements of
$\mathcal{B}_1^{'}$ not in this common subspace can be written, up to a
relative phase, as $\ket{\nu_{j'}^{'}}=\cos\theta_{j'}\ket{m_{j'}}
+\sin\theta_{j'}\ket{n_{j'}}$, where
$\ket{m_{j'}}\in\text{span}(\mathcal{B}_0)$,
$\ket{n_{j'}}\in\ker(\Pi_0)$, and $0<\theta_{j'}<\pi/2$.

If $\Pi_1$ induces a non-trivial isometry on
$\text{span}(\mathcal{B}_0)$, this implies that $\Pi_0$ and $\Pi_1$ do
not share any common non-trivial eigenvectors, i.e. $k=k'$, otherwise
the transition probabilities for states in the common eigenspace and
those with outside support would differ. Hence we can express
$\mathcal{B}_1^{'}=
\left\{\cos\theta\ket{m_{j'}}+\sin\theta\ket{n_{j'}}\right\}_{j'=1}^{k}$,
and the transition probability between the subspaces is
$t^2=\cos^2\theta$. Completing the resolution of the identity
specifies a measurement, and the complementary outcome to $\Pi_1$ given by
$\widetilde{\Pi}_1=\one-\Pi_1$ is also a projective isometry that
translates the subspace of $\Pi_0$ to one that is orthogonal to
$\Pi_1$. If we make a two-outcome measurement
$\{\Pi_1,\widetilde{\Pi}_1\}$, regardless of the resulting projection
the information originally in the subspace of $\Pi_0$ is preserved. We
now use this to construct a repeat-until-success protocol.


\begin{figure}
\includegraphics[width=\columnwidth]{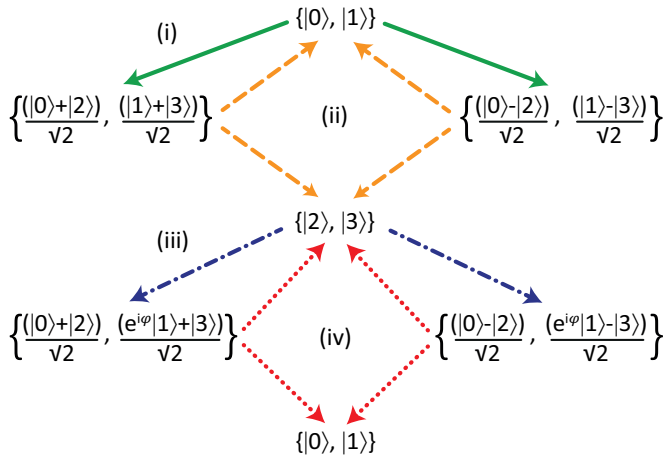}
\caption{(Color online) Qubit unitary projection sequence. Each
  two-outcome measurement is indicated by a pair of similar arrows
  between nodes representing the initial and final subspaces
  associated with degenerate projections. (i) The qubit is initially
  in the $\{\ket{0},\ket{1}\}$ subspace and the first measurement
  (solid green lines) projects equiprobably onto the
  $\{(\ket{0}+\ket{2})/\sqrt{2},(\ket{1}+\ket{3})/\sqrt{2}\}$ subspace
  or its complement,
  $\{(\ket{0}-\ket{2})/\sqrt{2},(\ket{1}-\ket{3})/\sqrt{2}\}$. (ii)
  The second measurement (dashed yellow) projects back onto the
  original $\{\ket{0},\ket{1}\}$ subspace or forwards onto the
  $\{\ket{2},\ket{3}\}$ subspace that acts as a ``checkoint''. In the
  former case, the qubit is unaltered and we begin again. (iii) For
  the $\{\ket{2},\ket{3}\}$ outcome, the next (third) measurement
  (dash-dotted blue) begins the return loop via a different
  subspace. (iv) The fourth measurement (dotted red) either projects
  back to the $\{\ket{2},\ket{3}\}$ checkpoint subspace or completes
  the loop returning the system to the initial subspace. The qubit
  will have undergone a unitary holonomy
  $U=e^{i\frac{\varphi}{2}}e^{-i\varphi\frac{\sigma_z}{2}}$, up to a
  $\pm 1$ global phase factor depending on the path taken through the
  graph. The rotation axis of $U$ can be chosen by redefinition of
  $\{\ket{0},\ket{1}\}$. As each transition is equally weighted
  $(p=\frac{1}{2})$, the mean time for traversing the graph top to
  bottom can be shown to be 8 steps.}
\label{fig:projtree}
\end{figure}

We illustrate the basic idea by implementing isometries on a qubit
initially embedded in the $\{\ket{0},\ket{1}\}$ subspace. We augment
the system by the direct sum of two additional levels,
$\{\ket{2},\ket{3}\}$ and define a measurement with two
$2$-dimensional degenerate projections with subspaces,
\begin{eqnarray}
\Pi_1&:&\{\cos\theta\ket{0}+e^{i\phi}\sin\theta\ket{2},
\cos\theta\ket{1}+e^{i\varphi}\sin\theta\ket{3}\}
\nonumber\\
\widetilde{\Pi}_1&:&\{\sin\theta\ket{0}-e^{i\phi}\cos\theta\ket{2},
\sin\theta\ket{1}-e^{i\varphi}\cos\theta\ket{3}\},
\end{eqnarray}
where the phases $\{\phi,\varphi\}$ are arbitrary. The measurement
probabilities are independent of the initial state and are
$P_1=\cos^2\theta$ and $P_{\tilde{1}}=\sin^2\theta$
respectively. Conditional on the result, we can choose different pairs
of degenerate projections to measure in the next round, each
subsequent measurement may depend on previous outcomes
resulting in a binary tree of projections~\cite{AO2008}. If at some
point in the sequence the resultant subspace returns to the original
one, a unitary holonomy would be the result. A randomly choosen
sequence of such measurements periodically revisiting the original
subspace can generate a desired unitary in an expected number of steps
polynomial in the approximation error~\cite{SO2013A,SO2013B,SO2014}.

Alternatively, it may be more efficient for the projection sequence to
trace out specific paths, Fig.~\ref{fig:projtree} demonstrates such a
sequence that implements a near deterministic qubit rotation. An
initial qubit state is translated between different subspaces
according to the directed graph structure until it returns to the
original subspace having had applied to it a unitary holonomy.
Due to measurement randomness, the measurements may need to be
repeated until a successful sequence of results is obtained. The
probability of failure decreases exponentially in the total number of
allowed steps and repeat-until-success protocols have been used to
good effect in gate synthesis~\cite{PS2013}.


As a concrete example, we implement
$U=e^{i\frac{\varphi}{2}}e^{-i\varphi\frac{\sigma_z}{2}}$ on an
initial state $\ket{\psi}=\alpha\ket{0}+\beta\ket{1}$ assuming that we
take the minimal traversal (no backtracking) down the right side of the
graph in Fig.~\ref{fig:projtree}. The first measurement takes us to
the subspace
$\{(\ket{0}-\ket{2})/\sqrt{2},(\ket{1}-\ket{3})/\sqrt{2}\}$ and the
system becomes
\begin{eqnarray}
  \ket{\psi_1}&=&\frac{
\left(
\frac{\ket{0}-\ket{2}}{\sqrt{2}}
\frac{\bra{0}-\bra{2}}{\sqrt{2}}
+
\frac{\ket{1}-\ket{3}}{\sqrt{2}}
\frac{\bra{1}-\bra{3}}{\sqrt{2}}
\right)}
{\sqrt{\frac{1}{2}}}\ket{\psi}\nonumber\\
&=&\alpha\frac{\ket{0}-\ket{2}}{\sqrt{2}}
  +\beta\frac{\ket{1}-\ket{3}}{\sqrt{2}},
\end{eqnarray}
where in the first line we have normalized the state by the square
root of the transition probability which is independent of
$\ket{\psi}$ as required by isometry. If the second projection does
not take us back up to the original subspace but down to the next
``checkpoint'' subspace $\{\ket{2},\ket{3}\}$ instead, the state is
now
\begin{equation}
  \ket{\psi_2}=-(\alpha\ket{2}+\beta\ket{3}).
\end{equation}
The third measurement begins the return loop but via a different
subspace, an outcome to the right results in
\begin{equation}
  \ket{\psi_3}=\alpha\frac{\ket{0}-\ket{2}}{\sqrt{2}}
  +\beta\frac{e^{i\varphi}\ket{1}-\ket{3}}{\sqrt{2}}.
\end{equation}
A final successful measurement completes the loop leaving the qubit
back in its original subspace,
\begin{equation}
  \ket{\psi_4}=\alpha\ket{0}+e^{i\varphi}\beta\ket{1}
=U\ket{\psi}
\end{equation}
as required. A simple calculation shows that holonomies corresponding
to different traversals only differ by a $\pm 1$ global phase factor.


\label{generalizedholonomy}

We can generalize the procedure in Fig.~\ref{fig:projtree} to induce a
$k$-dimensional unitary on an initial subspace spanned by
$\left\{\ket{j}\right\}_{j=1}^{k}$. We augment the Hilbert space with
an additional $k'=k$ levels $\left\{\ket{\bar{j}}\right\}_{j}^{k}$. We
now project onto subspaces spanned by (unnormalized) vectors
$\left\{\ket{j}+\ket{\bar{j}}\right\}$ and
$\left\{\ket{j}-\ket{\bar{j}}\right\}$ for the first measurement,
$\left\{\ket{j}\right\}$ and $\left\{\ket{\bar{j}}\right\}$ for the
second measurements, $\left\{e^{i\phi_j}\ket{j}+\ket{\bar{j}}\right\}$
and $\left\{e^{i\phi_j}\ket{j}-\ket{\bar{j}}\right\}$ for the third
measurements, and $\left\{\ket{j}\right\}$ and
$\left\{\ket{\bar{j}}\right\}$ for the fourth and final measurements
in the graph. The induced holonomy after a successful sequence of
projections is given by $U=\text{diag}\left(e^{i\phi_m}\right)$ in the
$\left\{\ket{j}\right\}$ basis, up to a $\pm 1$ global phase. The
graph structure is identical to that in Fig.~\ref{fig:projtree} with
the same transition probabilities and expected transit time of 8
steps.


In summary we have elucidated criteria and restrictions for inducing
isometries between subspaces by discrete projections, complementing
previous work exploring the Zeno regime~\cite{Burgarth2013,Lidar2013}
or formal aspects of projective
holonomies~\cite{AP1989,KAS2006,SKA2006}. In order to preserve
information during each projection, the dimensionality of the entire
space must be at least twice that of the initial subspace. The direct
and iterative holonomies coincide in this case~\cite{SKA2006}. Using a
cyclic sequence of projections we construct a repeat-until-success
protocol to implement a general unitary operation with an average of 8
measurements. If doubling the initial subspace dimension is not
possible, we also show that a single additional level is sufficient
for inducing a unitary holonomy. In this case, the trade-off is in the
success probability, though it can be increased with more projections
until we ultimately recover the Zeno effect in the infinite limit.

The required highly degenerate projections may be possible
experimentally. The proposal in~\cite{Oi2013} suggests a way of
performing infinitely degenerate projections on photon number in
cavity quantum electrodynamics with displacements and squeezing to
effectively modify the projection subspace. An intriguing possibility
in such infinite dimensional systems is the creation of additional
empty levels, as in the Hilbert Hotel
Paradox~\cite{HilbertHotel,POJ2014}, to act as ancillary dimensions as
required for stepwise isometries. This may require the development of
more non-Gaussian operations in order to perform the required
manipulations of the states to project onto different subspaces in
conjunction with the techniques outlined in~\cite{Oi2013}.

Comparing this work with measurement based quantum computation
(MBQC)~\cite{MBQC2003} and ancilla driven computation
(ADQC)~\cite{Twisted2009,ADQC2010,Twisted2012} which also employ
measurement to drive dynamics, the key differences are that in the
latter two schemes, a tensor product structure is assumed and
measurement is performed on subsystems, not
subspaces~\cite{AO2008,WY2006}. Our results are more general since a tensor
product space can be mapped to a direct sum, but not neccessarily the
converse. The minimal addition of a single qubit (e.g. in ADQC)
automatically doubles the dimensionality and this doubling is both
necessary and sufficient for unitary conditional evolution via
projections on the qubit. In some experimental implementations,
e.g. using position degrees of freedom~\cite{SOG2005}, it is
comparatively easy to increase dimensionality by the direct sum of
ancillary levels, rather than add subsystems and couple them to
perform entangling operations.

We finally note that near deterministic unitary holonomies require
that coherence is preserved at each step. The results
of~\cite{Oi2014b} and references therein show that it is impossible,
with unit probability, to ``unlearn'' information gained from a
measurement outcome whose Kraus operator does not have a flat
distribution of singular values. Hence this rules out the possibility
of measurement trees or graphs where all final cumulative results are
unitary but for which some of the intermediate effects are not
isometries. The two classes of protocol presented lie at the ends of
the spectrum, either preserving coherence at every step, or else only
one of the final outcomes is unitary with the rest collapsing the
state.

\begin{acknowledgments}
  DKLO thanks Si-Hui Tan, Kuldip Singh, John Jeffers,
  V{\'a}clav Poto{\v c}ek, Kerem Halil-Shah, and Johan {\AA}berg for
  comments and useful discussions, and is supported by Quantum
  Information Scotland (QUISCO).
\end{acknowledgments}

\end{document}